\documentclass[superscriptaddress,secnumarabic,prd,aps,showpacs,nofootinbib,showkeywords,eqsecnum]{revtex4-1}

\usepackage{graphicx,color,amsmath,amsxtra}
\usepackage{epsf}
\usepackage{amssymb}
\usepackage{enumerate}
\usepackage{hhline}
\usepackage{array}
\usepackage{tabularx}
\usepackage[unicode]{hyperref}
\usepackage{graphicx}                
\usepackage{epstopdf}

\begin{document}
\pagestyle{myheadings}

\title{Anisotropic power-law inflation for a model of two scalar and two vector fields}
\author{Tuan Q. Do }
\email{tuan.doquoc@phenikaa-uni.edu.vn}
\affiliation{Phenikaa Institute for Advanced Study, Phenikaa University, Hanoi 12116, Vietnam}
\affiliation{Faculty of Basic Sciences, Phenikaa University, Hanoi 12116, Vietnam}
\author{W. F. Kao}
\email{gore@mail.nctu.edu.tw}
\affiliation{Institute of Physics, National Chiao Tung University, Hsin Chu 30010, Taiwan, \\ and \\
Institute of Physics, National Yang Ming Chiao Tung University, Hsin Chu 30010, Taiwan
}
\date{\today}

\begin{abstract}
Inspired by an interesting counterexample to the cosmic no-hair conjecture found in a supergravity-motivated model recently, we propose a multi-field extension, in which two scalar fields are allowed to non-minimally couple to two vector fields, respectively. This model is shown to admit an exact Bianchi type I power-law solution. Furthermore, stability analysis based on the dynamical system method is performed to show that this anisotropic solution is indeed stable and attractive if both scalar fields are canonical. Nevertheless, if one of the two scalar fields is phantom then the corresponding anisotropic power-law inflation turns unstable as expected. 
\end{abstract}

%


\maketitle
\section{Introduction} \label{intro}
Cosmic inflation \cite{guth} has played a central paradigm in modern cosmology due to the fact that its predictions have been well confirmed by the cosmic microwave background (CMB) radiations probes such as the Wilkinson Microwave Anisotropy Probe (WMAP)  ~\cite{WMAP} and the  Planck ~\cite{Planck}. It is widely believed that a hypothetical scalar inflaton field is responsible for the inductance of the inflationary phase during the early universe \cite{Martin:2013tda}. Remarkably, the Starobinsky model, one of the original inflation models \cite{guth}, still remains as one of the most favorable models in light of the Planck observation ~\cite{Planck}. It is important to note that the cosmological principle, stating that our universe on large scales is simply homogeneous and isotropic Friedmann-Lemaitre-Robertson-Walker (FLRW) spacetime \cite{FLRW}, has been the key base of most inflationary models \cite{Martin:2013tda}. However, testing the validity of the cosmological principle is not a straightforward task \cite{cosmological-principle}.

Recently, observations of CMB anomalies, such as the hemispherical asymmetry and the cold spot observed by WMAP and Planck, have been one of the great challenges of the standard inflationary models based on the cosmological principle \cite{Schwarz:2015cma}. 
Of course, there have been a number of mechanisms proposed to explain the origin of these anomalies \cite{Schwarz:2015cma}. For example, there have been claims  in Ref. \cite{Hanson:2010gu} that the CMB statistical anisotropy could be instrumental rather than cosmological. In particular, these investigations have shown that the asymmetric beams could be the origin of the CMB statistical anisotropy.  
This resolution has, however, been tested independently by other people. 
Indeed, it was pointed out that the asymmetric beams seem to be unimportant \cite{Groeneboom:2009cb}.

It turns out that investigating the origin of these exotic features might lead to a small deviation from the cosmological principle. 
One of the possible deviations is to introduce the Bianchi spacetimes, which are homogeneous but anisotropic metrics, instead of the FLRW one in order to describe the early universe during the inflationary phase \cite{bianchi}. 
Interestingly, some theoretical predictions of the Bianchi type inflationary universe have already been derived in Ref. \cite{Pitrou:2008gk}. 
If the early universe was anisotropic initially, it is important to answer the question whether the universe is still anisotropic or not.
Of course, this question is not easy to answer either by theoretical derivation or by direct observation. 
On the observational side, it is worth noting that a recent study in Ref. \cite{Colin:2018ghy} might provide us with an important hint to this question. In particular, it has been claimed that the present universe might be anisotropic rather than isotropic. 
On the theoretical side, there is an important hint provided by the so-called cosmic no-hair conjecture proposed by Hawking and his colleagues a few decades ago \cite{GH}. 
The cosmic no-hair conjecture states that the late time universe should be inevitably homogeneous and isotropic consistent with the cosmological principle regardless of the early state of the universe.
If this conjecture is not valid, the observation shown in Ref. \cite{Colin:2018ghy} might be a relevant evidence. 
It is worth noting that the cosmic no-hair conjecture, if correct, should be valid locally, i.e., inside of the future event horizon, as shown firstly by Starobinsky \cite{Starobinsky:1982mr} and then by other people \cite{Barrow:1984zz}. It turns out that a number of huge efforts have been paid to prove this conjecture since the first partial proof by Wald for the Bianchi spacetimes \cite{wald,Barrow:1987ia,inhomogeneous,Carroll:2017kjo}. However, a complete proof for this conjecture has still remained a great challenge.

Along with these  proofs, there have been claims that the cosmic no-hair conjecture could no longer be valid \cite{barrow06,kaloper,galileon,kao09,MW0,MW}. Nevertheless, stability analysis done in Ref. \cite{kao09} have pointed out that some of the claimed counterexamples turn out to be unstable during an inflationary phase, consistent with the prediction of the conjecture. On the other hand, there existed a very truly counterexample to the conjecture found in the supergravity-motivated model  by Kanno-Soda-Watanabe (KSW) \cite{MW0,MW}. 
In particular, the KSW model introduces saclar-vector coupling term $f^2(\phi)F_{\mu\nu}F^{\mu\nu}$ which breaks the conformal invariance. 
As a result, the existence of the non-trivial function $f(\phi)$ prevents the vector field from an exponential dilution during the inflationary phase. 
Hence, the existence of the non-vanishing vector field will lead to small spatial anisotropies during the inflationary phase. 
It turns out that the KSW model does admit the Bianchi type I spacetime as its stable and attractive inflationary solution \cite{MW0,MW}. Consequently, this model has been investigated extensively \cite{extensions,WFK,Do:2017rva,Fujita:2018zbr,multi-vector-1,multi-vector-2,ghost-condensed,non-canonical,Do:2018zac,data,Imprint1,Imprint2,Imprint3,Do:2020ler,gws,SD}. For example, non-trivial extensions with multi scalar fields coupled to one vector field have been proposed in Refs. \cite{WFK,Do:2017rva,Fujita:2018zbr}; while other non-trivial extensions with multi vector fields coupled to one scalar field have been studied in Refs. \cite{multi-vector-1,multi-vector-2}. More interestingly, non-canonical scenarios, in which a non-canonical scalar field is coupled to a vector field, have been investigated in Refs. \cite{ghost-condensed,non-canonical}. Additionally, a higher dimensional version of the KSW model has been examined in Ref. \cite{Do:2018zac}. It turns out that many counterexamples to the cosmic no-hair conjecture have been found not only in these extensions but also in other models shown in Ref. \cite{extensions}. This result indicates that the unusual coupling $f^2(\phi)F_{\mu\nu}F^{\mu\nu}$ does play an interesting role in the validity of the cosmic no-hair conjecture. In order to compare with the future CMB observations, expected to have higher sensitivity, the CMB imprints of the anisotropic inflation \cite{ACW} have been investigated systematically in Refs. \cite{Imprint1,Imprint2,Imprint3,Do:2020ler}. Additionally, primordial gravitational waves within the anisotropic inflation have also been studied in Ref. \cite{gws}. Many other cosmological features of the KSW anisotropic inflation can also be found in Ref. \cite{SD}.

It is worth noting that the conformal invariance of the electromagnetic field has been believed  to be broken in order to generate non-trivial magnetic fields \cite{Turner:1987bw}. 
Indeed, the conformal-violating gauge coupling, $\exp[\phi] F_{\mu\nu}F^{\mu\nu}$, similar to the KSW model \cite{MW}, has been proposed by Ratra in Ref. \cite{Ratra:1991bn} as a natural origin of large-scale galactic electromagnetic fields in the present universe. 
Hence, there is a close relation between the existence of spatial anisotropies during the inflationary phase and the late time large-scale galactic electromagnetic fields. In other words, the existence of the late time large-scale galactic electromagnetic fields might be a smoking gun for the anisotropic inflationary universe. 
In addition, there are other types of conformal-violating Maxwell coupling that also lead to anisotropic inflation\cite{Holland:2017cza,Adak:2016led}.

Motivated by these observations, we would like to propose in this paper a non-trivial extension of the KSW model with multi scalar fields and multi vector fields. In particular, we will start by focusing on the model with two scalar fields non-minimally coupled to two vector fields similar to the KSW model. 
Hence, this model can be regarded as a non-trivial combination of the two-scalar-field models \cite{WFK} and the multi vector fields models \cite{multi-vector-1,multi-vector-2}. As a result, we are able to obtain a set of Bianchi type I power-law inflationary solutions to this model. 
Furthermore, we will also show that if both scalar fields are canonical then the corresponding solution will be stable and attractive. This solution, therefore, acts as a new counterexample to the cosmic no-hair conjecture. On the other hand, if one of the two scalar fields is a phantom field \cite{Caldwell:1999ew,Guo:2004fq,Chimento:2008ws,Arefeva:2009tkq,Cai:2009zp} with negative-definite kinetic energy, then the corresponding solution will turn unstable similar to our previous investigations \cite{WFK}.

In summary, this paper will be organized as follows: (i) A brief introduction has been presented in Sec. \ref{intro}. (ii) A new model with two scalar fields coupled to two vector fields will be introduced in Sec. \ref{sec1}. (iii) Anisotropic power-law solution will be solved in Sec.  \ref{sec2}. (iv) Then, the stability of the new solution will be analyzed by the dynamical system method in Sec. \ref{sec3}. (v) Finally, concluding remarks will be drawn in Sec. \ref{sec4}.
\section{The model}  \label{sec1}
In this paper, we would like to propose a multi-field extension of the KSW model, in which two scalar fields are allowed to couple to two vector fields, respectively, as follows  
 \begin{align} \label{action}
S = \int d^4 x\sqrt{-g} \left[ \frac{1}{2}R -\frac{1}{2}\partial_\mu \phi \partial^\mu \phi -\frac{\omega}{2}\partial_\mu \psi \partial^\mu \psi -V_1(\phi) -V_2(\psi) - \frac{f_1^2(\phi)}{4}  F_{\mu\nu}F^{\mu\nu} - \frac{f_2^2(\psi)}{4}  {\cal F}_{\mu\nu} {\cal F}^{\mu\nu} \right] ,
\end{align}
where the reduced Planck mass $M_p$ has been set to be one for convenience. In addition, $\phi$ and $\psi$ are scalar fields, while $F_{\mu\nu} \equiv \partial_\mu A_\nu -\partial_\nu A_\mu$ and ${\cal F}_{\mu\nu} \equiv \partial_\mu {\cal A}_\nu -\partial_\nu {\cal A}_\mu$ are the field strength of the vector fields $A_\mu$ and ${\cal A}_\mu$, respectively. Note that $\psi$ will be canonical or phantom \cite{Caldwell:1999ew,Guo:2004fq,Chimento:2008ws,Arefeva:2009tkq,Cai:2009zp} if $\omega$ is equal to $1$ or $-1$, respectively. 

Before going to discuss this model in details, we would like to mention that some multi-field extensions of the KSW model have been proposed. In particular, two scalar fields models, in which one scalar field is phantom  and both of these scalar fields coupled to a vector field, have been proposed in Refs. \cite{WFK,Do:2017rva}. As a result, the inclusion of the phantom scalar field leads to an instability of the anisotropic inflation, making the corresponding model consistent with the cosmic no-hair conjecture \cite{WFK,Do:2017rva}. On the other hand, multi vector fields models, in which one canonical scalar field is coupled to multi vector fields, have been studied in Refs. \cite{multi-vector-1,multi-vector-2,Do:2018zac}. Hence, our present model turns out to be more general than these models since it deals with multi scalar fields coupled to multi vector ones. 

As a result, the corresponding Einstein field equation of this model is derived to be
\begin{align} \label{Einstein-field}
& R_{\mu\nu} -\frac{1}{2}Rg_{\mu\nu } -\partial_\mu \phi \partial_\nu \phi  -\omega \partial_\mu \psi \partial_\nu \psi +g_{\mu\nu} \left[\frac{1}{2}\partial_\sigma \phi \partial^\sigma \phi +\frac{\omega}{2}\partial_\sigma \psi \partial^\sigma \psi +V_1+V_2 +\frac{1}{4} \left(f_1^2 F^2+  f_2^2{\cal F}^2 \right) \right] \nonumber\\
&-f_1^2 F_{\mu\gamma}F_\nu{}^{\gamma} -f_2^2 {\cal F}_{\mu\gamma} {\cal F}_\nu{}^{\gamma}=0.
\end{align}
In addition, the corresponding field equations of these two vector fields, $A_\mu$ and ${\cal A}_\mu$, are given by 
\begin{align} \label{vector-field-1}
\partial_\mu \left[\sqrt{-g}  f_1^2 F^{\mu\nu}  \right] &=0,\\
\label{vector-field-2}
\partial_\mu \left[\sqrt{-g}  f_2^2 {\cal F}^{\mu\nu} \right] &=0,
\end{align}
along with that of the scalar field $\phi$,
\begin{align} \label{phi-scalar-field}
\square \phi -\partial_\phi V_1-\frac{1}{2}  f_1 \left(\partial_\phi f_1\right) F^2 &=0,\\
\label{psi-scalar-field}
\omega \square \psi -\partial_\psi V_2-\frac{1}{2}f_2 \left(\partial_\psi f_2\right) {\cal F}^2 &=0,
\end{align}
where $\partial_\phi \equiv \partial/\partial \phi $, $\partial_\psi \equiv \partial/\partial \psi $, and $\square \equiv \frac{1}{\sqrt{-g}} \partial_\mu \left(\sqrt{-g} \partial^\mu \right)$. 
In this paper, our purpose is to figure out anisotropic power-law solutions to this model. Hence, we will work with the Bianchi type I metric, which is the simplest homogeneous but anisotropic spacetime, whose form is given by \cite{MW0,MW}
\begin{equation} \label{metric}
ds^2 =-dt^2 +\exp\left[ 2\alpha(t) -4\sigma(t) \right] dx^2 +\exp\left[ 2\alpha(t) +2\sigma(t) \right] \left(dy^2+dz^2 \right),
\end{equation}
where $\sigma(t)$ is a deviation from the spatial isotropy governed by $\alpha(t)$. This means that $\sigma(t) \ll \alpha(t)$ is a sufficient condition during an inflationary phase.  In addition, the vector fields, $A_\mu$  and ${\cal A}_\mu$, are chosen as  $A_\mu   = \left( {0,A_x \left( t \right),0,0} \right)$ and  ${\cal A}_\mu   = \left( {0,{\cal A}_x \left( t \right),0,0} \right)$ in order to be compatible with the Bianchi metric having the $y-z$ rotational symmetry as proposed in Eq. \eqref{metric}. The last ingredients of the model, i.e., the scalar fields, are assumed to be homogeneous, i.e., $\phi =\phi(t)$ and $\psi=\psi(t)$. 

As a result, the vector field equations, Eqs. \eqref{vector-field-1} and \eqref{vector-field-2}, can be solved to give non-trivial solutions such as
\begin{align}
\dot A_x & =p_A f_1^{-2}  \exp[-\alpha-4\sigma],\\
{\dot {\cal A}}_x &=q_A f_2^{-2} \exp[-\alpha-4\sigma],
\end{align}
where $p_A$ and $q_A$ are integration constants. Consequently,  the field equations \eqref{Einstein-field}, \eqref{phi-scalar-field}, and \eqref{psi-scalar-field} now turn out to be 
\begin{align} \label{field-equation-1}
\dot\alpha^2 &= \dot\sigma^2 +\frac{1}{3} \left[ \frac{\dot\phi^2}{2}+\omega \frac{\dot\psi^2}{2}+V_1 +V_2 +\frac{1}{2} \left(p_A^2 f_1^{-2}+q_A^2 f_2^{-2}  \right)\exp[-4\alpha-4\sigma] \right],\\
\label{field-equation-2}
\ddot\alpha&=-3\dot\alpha^2 +V_1+V_2 +\frac{1}{6} \left(p_A^2 f_1^{-2}+q_A^2 f_2^{-2}  \right) \exp[-4\alpha-4\sigma],\\
\label{field-equation-3}
\ddot\sigma&=-3\dot\alpha \dot\sigma +\frac{1}{3} \left(p_A^2 f_1^{-2}+q_A^2 f_2^{-2}  \right) \exp[-4\alpha-4\sigma],\\
\label{field-equation-4}
\ddot\phi&=-3\dot\alpha \dot\phi -\partial_\phi V_1 +  p_A^2 f_1^{-3}(\partial_\phi f_1)\exp[-4\alpha-4\sigma],\\
\label{field-equation-5}
\ddot\psi&=-3\dot\alpha \dot\psi - \frac{1}{\omega} \left[ \partial_\psi V_2 - q_A^2 f_2^{-3} (\partial_\psi f_2) \exp[-4\alpha-4\sigma] \right] .
\end{align} 
It is noted that Eq. \eqref{field-equation-1} is nothing but the Friedmann equation, which plays as a constraint field equation. In addition, Eqs. \eqref{field-equation-2} and  \eqref{field-equation-3} act as evolution equations of the spatial isotropy $\alpha$ and anisotropy $\sigma$, respectively. 
\section{Anisotropic power-law solution} \label{sec2}
Now, we would like to investigate whether anisotropic power-law inflation appears within this model. This investigation follows the previous studies presented in Refs. \cite{MW,WFK}. It is noted that an isotropic power-law inflation was found quite long time ago, e.g., see Ref. \cite{Abbott:1984fp}.  In particular, we will consider the following ansatz such as \cite{MW,WFK}
\begin{equation}
\alpha (t)= \zeta \log t; ~\sigma(t) = \eta \log t;~ \phi(t) = \xi_1 \log t +\phi_0;~\psi(t) = \xi_2 \log t +\psi_0
\end{equation}
along with the compatible exponential potential and coupling functions such as
\begin{align}
V_1(\phi)&=V_{01} \exp[\lambda_1 \phi],\\
V_2(\psi)&=V_{02} \exp[\lambda_2 \psi],\\
f_1(\phi)&=f_{01}\exp[\rho_1 \phi],\\
f_2(\psi)&=f_{02}\exp[\rho_2 \psi],
\end{align}
here $\phi_0$, $\psi_0$, $\xi_i$, $V_{0i}$, $f_{0i}$, $\lambda_i$, and $\rho_i$ are all non-vanishing parameters. In addition, $\zeta$ and $\eta$ are parameters characterizing the power-law  expansion of spacetimes as
\begin{equation}
\exp[2\alpha-4\sigma]=t^{2\zeta-4\eta}; ~\exp[2\alpha+2\sigma] =t^{2\zeta+2\eta}.
\end{equation}
As a result, the value of $\zeta$ and $\eta$ will be determined after solving the field equations. It is noted that $\zeta-2\eta>0$ and $\zeta+\eta>0$ are two sufficient constraints for expanding universe. However, if an expanding universe become an inflationary universe, these constraints should be modified to be $\zeta-2\eta \gg 1$ and $\zeta+\eta\gg 1$ \cite{MW,WFK}.

As a result, the corresponding set of algebraic equations derived from the field equations \eqref{field-equation-1}, \eqref{field-equation-2}, \eqref{field-equation-3} and \eqref{field-equation-4} turn out to be
\begin{align}
\label{algebraic-1}
\zeta^2 &= \eta^2 +\frac{1}{3} \left[\frac{\xi_1^2}{2}+\omega \frac{\xi_2^2}{2}+ u_1+u_2 +\frac{1}{2} \left(v_1 +v_2 \right) \right],\\
\label{algebraic-2}
-\zeta&= -3\zeta^2 +u_1+u_2 +\frac{1}{6} \left(v_1+v_2 \right),\\
\label{algebraic-3}
-\eta&= -3\zeta \eta +\frac{1}{3} \left(v_1+v_2  \right),\\
\label{algebraic-4}
-\xi_1 &= -3\zeta \xi_1 -\lambda_1 u_1 + \rho_1  v_1,\\
\label{algebraic-5}
-\xi_2 &= -3\zeta \xi_2 - \frac{1}{\omega} \left(\lambda_2 u_2 - \rho_2  v_2\right).
\end{align}
It is noted that in order to have these algebraic equations, we have imposed the corresponding constraint equations,
\begin{align}
\label{constraint-1}
\lambda_1 \xi_1 &= -2,\\
\label{constraint-2}
\lambda_2 \xi_2 &= -2,\\
\label{constraint-3}
2\zeta+2\eta+\rho_1 \xi_1 & =1,\\
\label{constraint-4}
2\zeta+2\eta+\rho_2 \xi_2 & =1,
\end{align}
which lead all terms in the field equations to be functions of $t^{-2}$. It is also noted that we have defined additional variables $u_i$ and $v_i$ ($i=1-2$) as
\begin{align}
u_1=&V_{01} \exp \left[\lambda_1 \phi_0 \right],\\
u_2=&V_{02} \exp \left[\lambda_2 \psi_0 \right],\\
v_1= & p_A^2 f_{01}^{-2} \exp[-2\rho_1\phi_0],\\
v_2= & q_A^2 f_{02}^{-2} \exp[-2\rho_2\psi_0].
\end{align}
It turns out from the above constraints  that
\begin{align} \label{constraint-5}
\lambda_1 \xi_1& =\lambda_2 \xi_2,\nonumber\\
\rho_1 \xi_1  &=\rho_2 \xi_2,
\end{align}
which imply
\begin{align} \label{constraint-kappa-1}
\frac{\rho_1}{\lambda_1} =\frac{\rho_2}{\lambda_2} &= \kappa_1,\\
\label{constraint-kappa-2}
\lambda_1 \rho_2 =\lambda_2\rho_1 &= \kappa_2,
\end{align}
where $\kappa_1$ and $\kappa_2$ are additional constants. Up to now, we have six variables, $\zeta$, $\eta$, $u_1$, $u_2$, $v_1$, and $v_2$ needed to be solved from the five independent algebraic equations \eqref{algebraic-2}, \eqref{algebraic-3}, \eqref{algebraic-4}, \eqref{algebraic-5},  and \eqref{constraint-3} (or \eqref{constraint-4}). Hence, explicit analytical values of these variables cannot be solved altogether from these equations. However, we will show that it is possible to figure out explicit values of $\zeta$ and $\eta$ from the above algebraic equations, even when that of $u_i$ and $v_i$ remain unsolved.

 Note that the obtained solutions should satisfy the constraint equation \eqref{algebraic-1}, which is derived from the Friedmann equation \eqref{field-equation-1}. As a result, we have from Eqs. \eqref{algebraic-4} and \eqref{algebraic-5} that
\begin{equation} \label{algebraic-6}
\left(3\zeta-1 \right) \left( \rho_2\xi_1 +\omega  \rho_1 \xi_2\right) = -\kappa_2 \left(u_1+u_2 \right) +\rho_1 \rho_2 \left(v_1+v_2 \right).
\end{equation}
As a result, by setting two additional variables
\begin{align}
u&=u_1+u_2, \\
v&=v_1+v_2,
\end{align}
we are able to figure out the value of $\zeta$ and $\eta$. Indeed, $u$ and $v$ can be solved from two Eqs. \eqref{algebraic-2} and \eqref{algebraic-3} as
\begin{align}
u&= \zeta \left(3\zeta-1\right)-\frac{v}{6},\\
v &= 3\eta \left(3\zeta-1\right).
\end{align}
Additionally, $\eta$ can be figured out from Eq. \eqref{constraint-3} (or \eqref{constraint-4}) to be
\begin{equation}
\eta =-\zeta+ \kappa_1 +\frac{1}{2}.
\end{equation}
As a result, plugging these solutions into Eq. \eqref{algebraic-6} with the help of the constraint \eqref{constraint-5} leads to an equation of $\zeta$,
\begin{equation} \label{equation-of-zeta-1}
\left(3\zeta-1\right) \left[ 6\lambda_1 \lambda_2 \left(\kappa_2 +2\rho_1 \rho_2\right) \zeta -  \lambda_1\lambda_2 \left(2\kappa_1+1\right) \left(\kappa_2+6\rho_1\rho_2 \right) -8\left(\omega \lambda_1 \rho_1+\lambda_2\rho_2 \right)\right]=0.
\end{equation}
Noting that the Friedmann equation \eqref{algebraic-1} can also be reduced to another equation of $\zeta$ as 
\begin{equation} \label{equation-of-zeta-2}
6\lambda_1^2 \lambda_2^2 \left(2\kappa_1 +1\right) \zeta - \lambda_1^2 \lambda_2^2 \left(12 \kappa_1^2 +8\kappa_1 +1\right) - 8 \left(\omega \lambda_1^2+\lambda_2^2 \right)=0.
\end{equation}
Ignoring a trivial solution, $\zeta=1/3$, which leads to an isotropic universe with $\eta=0$, we obtain a non-trivial solution of $\zeta$ from the above equation \eqref{equation-of-zeta-1},
\begin{equation} \label{solution-zeta}
\zeta = \frac{ \lambda_1\lambda_2 \left(2\kappa_1+1\right) \left(\kappa_2+6\rho_1\rho_2 \right) + 8\left(\omega \lambda_1 \rho_1+\lambda_2\rho_2 \right)}{6\lambda_1 \lambda_2 \left(\kappa_2 +2\rho_1 \rho_2\right) }.
\end{equation} 
It is straightforward to verify that this non-trivial solution does satisfy the constraint equation \eqref{equation-of-zeta-2} with noting the constraints shown in Eqs. \eqref{constraint-kappa-1} and \eqref{constraint-kappa-2}. Another quick cross-check is setting $\lambda_2 =\lambda_1=\lambda$, $\rho_2=\rho_1=\rho$, and $\omega=+1$, which corresponds to the KSW model with double identical scalar and vector fields. As a results, this solution will be reduced in this case to
\begin{equation}
\zeta = \frac{\lambda^2 +8\lambda \rho +12\rho^2+16}{6\lambda^2 \left(\lambda+2\rho\right)},
\end{equation}
which is consistent with the solution obtained in Ref. \cite{MW} for one scalar coupled to one vector field.
In other words, the solution shown in Eq. \eqref{solution-zeta} is  exactly our desired solution. Hence, the corresponding $\eta$ can now be defined to be
\begin{equation} \label{solution-eta}
\eta =\frac{\kappa_2 \lambda_1 \lambda_2 \left(2\kappa_1 +1\right) -4 \left(\omega \lambda_1 \rho_1 +\lambda_2 \rho_2 \right)}{3\lambda_1 \lambda_2 \left(\kappa_2 +2\rho_1 \rho_2\right)}.
\end{equation}
Now, we would like to see whether these solutions represent anisotropic inflationary universe. It turns out that if $\rho_i \gg \lambda_i$ then $\kappa_1 \gg 1$ and $\kappa_2 \ll \rho_1 \rho_2$. Consequently, 
\begin{align}
\zeta &\simeq \kappa_1 \gg 1 ,\\
\eta & \simeq \frac{1}{3},\\
u&\simeq 3\kappa_1^2,\\
v&\simeq 3 \kappa_1. 
\end{align}
This result confirms our expectation that the present model does admit an anisotropic power-law inflation with a small spatial hair.
More interestingly, it is straightforward to verify that we always have the result $\zeta \simeq \kappa_1$ for anisotropic power-law solutions, regardless of the nature of scalar fields as well as the number of scalar and vector fields counted in the KSW model. The reason is based on the fact that we can always define $u$ and $v$ as
\begin{equation}
u= \sum_{i=1}^n u_i; ~v= \sum_{i=1}^n v_i,
\end{equation}
and therefore we can always find out $\zeta$, whose leading term is nothing but $\kappa_1 \equiv \rho_i/\lambda_i=\rho_k/\lambda_k$. However, it appears that the $\zeta$ solution, $\zeta \simeq \kappa_1$,  is a half of that obtained in the two-scalar-field models \cite{WFK,Do:2017rva}; while the $\eta$ solution, $\eta \simeq 1/3$, remains the same, assuming $\rho_1/\lambda_1 =\rho_2/\lambda_2 \gg 1$.
\section{Stability analysis} \label{sec3}
In this section, we would like to investigate whether the obtained anisotropic power-law solution is attractive during the inflationary phase, following the previous investigations \cite{MW,Do:2017rva,non-canonical}. In order to do this task, we will transform the field equations into the corresponding autonomous equations of dynamical system. In particular, we will define dynamical variables as
\begin{align}
X&= \frac{\dot\sigma}{\dot\alpha};~Y_1 =\frac{\dot\phi}{\dot\alpha};~Y_2=\frac{\dot\psi}{\dot\alpha},\\
Z_1&=\frac{p_A f_1^{-1}}{\dot\alpha}\exp[-2\alpha-2\sigma],\\
Z_2&=\frac{q_A f_2^{-1}}{\dot\alpha}\exp[-2\alpha-2\sigma],\\
W_1& =\frac{\sqrt{V_1}}{\dot\alpha};~W_2 =\frac{\sqrt{V_2}}{\dot\alpha},
\end{align}
where $W_1$ and $W_2$ are auxiliary dynamical variables \cite{Do:2017rva,Guo:2004fq}. Thanks to these definitions, we are now able to define the following results,
\begin{align}
\frac{dX}{d\alpha} &=\frac{\ddot\sigma}{\dot\alpha^2} -\frac{\ddot\alpha}{\dot\alpha^2}X,\\
\frac{dY_1}{d\alpha}&=\frac{\ddot\phi}{\dot\alpha^2}-\frac{\ddot\alpha}{\dot\alpha^2}Y_1,\\
\frac{dY_2}{d\alpha}&=\frac{\ddot\psi}{\dot\alpha^2}-\frac{\ddot\alpha}{\dot\alpha^2}Y_2,\\
\frac{dZ_1}{d\alpha}&= - \left[2(X+1)+ \rho_1 Y_1\right] Z_1 -\frac{\ddot\alpha}{\dot\alpha^2}Z_1,\\
\frac{dZ_2}{d\alpha}&= - \left[2(X+1)+ \rho_2 Y_2\right] Z_2 -\frac{\ddot\alpha}{\dot\alpha^2}Z_2,\\
\frac{dW_1}{d\alpha}&= \left(\frac{\lambda_1}{2}Y_1 -\frac{\ddot\alpha}{\dot\alpha^2} \right)W_1,\\
\frac{dW_2}{d\alpha}&= \left(\frac{\lambda_2}{2}Y_2 -\frac{\ddot\alpha}{\dot\alpha^2} \right)W_2,
\end{align}
where $\alpha$ acts as a new time coordinate, which is related to the cosmic time $t$ as $d\alpha=\dot\alpha dt$. Thanks to the field equations \eqref{field-equation-2}, \eqref{field-equation-3}, \eqref{field-equation-4}, and \eqref{field-equation-5}, we are able to have the following dynamical system involving autonomous equations defined as
\begin{align}\label{autonomous-1}
\frac{dX}{d\alpha} &= X\left[3\left(X^2-1\right) +\frac{1}{2}\left(Y_1^2 +\omega Y_2^2 \right) +\frac{1}{3}\left(Z_1^2+Z_2^2 \right)\right] +\frac{1}{3}\left(Z_1^2+Z_2^2 \right),\\
\label{autonomous-2}
\frac{dY_1}{d\alpha} &=  Y_1 \left[ 3\left(X^2-1\right)+\frac{1}{2} \left(Y_1^2 +\omega Y_2^2 \right)+\frac{1}{3} \left(Z_1^2 +Z_2^2 \right)\right] +\rho_1 Z_1^2-\lambda_1 W_1^2 ,\\
\label{autonomous-3}
\frac{dY_2}{d\alpha} &=  Y_2 \left[ 3\left(X^2-1\right)+\frac{1}{2} \left(Y_1^2 +\omega Y_2^2 \right)+\frac{1}{3} \left(Z_1^2 +Z_2^2 \right)\right] +\frac{\rho_2}{\omega} Z_2^2- \frac{\lambda_2}{\omega} W_2^2 ,\\
\label{autonomous-4}
\frac{dZ_1}{d\alpha}&= Z_1 \left[ 3\left(X^2-1\right)+\frac{1}{2} \left(Y_1^2 +\omega Y_2^2 \right)+\frac{1}{3}\left(Z_1^2 +Z_2^2 \right)-2X-\rho_1 Y_1 +1 \right],\\
\label{autonomous-5}
\frac{dZ_2}{d\alpha}&= Z_2 \left[ 3\left(X^2-1\right)+\frac{1}{2} \left(Y_1^2 +\omega Y_2^2 \right)+\frac{1}{3}\left(Z_1^2 +Z_2^2 \right)-2X-\rho_2 Y_2 +1 \right],\\
\label{autonomous-6}
\frac{dW_1}{d\alpha}&= W_1 \left[3X^2 +\frac{1}{2} \left(Y_1^2 +\omega Y_2^2 \right)+\frac{1}{3}\left(Z_1^2 +Z_2^2 \right) +\frac{\lambda_1}{2}Y_1 \right],\\
\label{autonomous-7}
\frac{dW_2}{d\alpha}&= W_2 \left[3X^2 +\frac{1}{2} \left(Y_1^2 +\omega Y_2^2 \right)+\frac{1}{3}\left(Z_1^2 +Z_2^2 \right) +\frac{\lambda_2}{2}Y_2 \right].
\end{align}
It is noted that the Friedmann constraint equation \eqref{field-equation-1}, which can be rewritten as
\begin{equation} \label{autonomous-8}
W_1^2 +W_2^2 =-3\left(X^2-1\right)-\frac{1}{2}\left(Y_1^2 +\omega Y_2^2 \right)-\frac{1}{2}\left(Z_1^2 +Z_2^2\right),
\end{equation}
has been used in order to derive the above autonomous equations. Now, we would like to figure out anisotropic fixed points with $X\neq 0$ to this dynamical system. Mathematically, fixed points, both isotropic and anisotropic, are solutions of the following set of equations, 
\begin{equation}
\frac{dX}{d\alpha}=\frac{dY_1}{d\alpha} =\frac{dY_2}{d\alpha} =\frac{dZ_1}{d\alpha}=\frac{dZ_2}{d\alpha}  =\frac{dW_1}{d\alpha} =\frac{dW_2}{d\alpha} =0.
\end{equation}
As a result, we have from two equations, 
 ${dW_1}/{d\alpha}={dW_2}/{d\alpha}=0$,  that
\begin{align} \label{dyn-constraint-1}
\lambda_1 Y_1 &= \lambda_2 Y_2  ,\\
\label{dyn-constraint-2}
3X^2 +\frac{1}{2} \left(Y_1^2 +\omega Y_2^2 \right)+\frac{1}{3}Z^2&=- \frac{\lambda_1}{2}Y_1,
\end{align}
provided  another requirement  that $W_1 \neq 0$ and $W_2\neq0$. Hence, it appears that
\begin{equation} \label{relation-Y}
Y_2 =\frac{\lambda_1}{\lambda_2}Y_1.
\end{equation}
Note that  $Z$ is additional variable introduced as
\begin{equation} \label{definition-of-Z}
 Z^2 =Z_1^2 +Z_2^2,
\end{equation}
for convenience.
 In addition, two equations, ${dZ_1}/{d\alpha}={dZ_2}/{d\alpha}=0$, imply, with the help of Eq. \eqref{dyn-constraint-2}, that
\begin{align}
\rho_1 Y_1 &=\rho_2 Y_2,\\
\label{equation-X-Y-Z-1}
2X+\left(\frac{\lambda_1}{2}+\rho_1\right) Y_1 +2 &=0,
\end{align}
provided  a requirement that $Z_1 \neq 0$ and $Z_2\neq0$. Hence, it is straightforward to recover the constraints shown in Eqs. \eqref{constraint-kappa-1} and \eqref{constraint-kappa-2} needed for the existence of the above anisotropic power-law solution. Hence, Eq. \eqref{equation-X-Y-Z-1} can be reduced to
\begin{equation} \label{equation-X-Y-Z-2}
2X+\left(\frac{\lambda_1}{2}+\lambda_1 \kappa_1 \right) Y_1 +2 =0.
\end{equation}
 It is noted that the equation ${dX}/{d\alpha}=0$ implies 
\begin{equation}\label{equation-X-Y-Z-3}
X\left( \frac{\lambda_1}{2}Y_1+ 3 \right) -\frac{1}{3}Z^2 =0.
\end{equation}
Finally, combining both equations, $dY_1/d\alpha=0$ and $dY_2/d\alpha=0$, imply  that
\begin{equation}\label{equation-X-Y-Z-4}
-\left(\frac{\lambda_1}{2}Y_1 +3 \right)\left[ \left(\rho_2 +\omega \rho_1 \frac{\lambda_1}{\lambda_2} \right)Y_1 +\kappa_2 \right]+\left( \rho_1 \rho_2+\frac{\kappa_2}{6} \right)Z^2=0
\end{equation} 
 with the help of equations \eqref{autonomous-8} and \eqref{dyn-constraint-2}.

Up to now, we have obtained three equations \eqref{equation-X-Y-Z-2}, \eqref{equation-X-Y-Z-3}, and \eqref{equation-X-Y-Z-4} for three variables $X$, $Y_1$, and $Z^2$. As a result, solving these equations will give us non-trivial solutions 
\begin{align} \label{solution-X}
X=~& \frac{2\left[ \kappa_2 \lambda_1 \lambda_2 \left(2\kappa_1+1\right) -4\left(\omega \lambda_1 \rho_1 +\lambda_2 \rho_2 \right) \right]}{\lambda_1\lambda_2 \left(2\kappa_1+1\right) \left(\kappa_2+6\rho_1\rho_2 \right) + 8\left(\omega \lambda_1 \rho_1+\lambda_2\rho_2 \right)},\\
Y_1 =~&\frac{-12\lambda_2 \left(\kappa_2 +2\rho_1 \rho_2 \right)}{\lambda_1\lambda_2 \left(2\kappa_1+1\right) \left(\kappa_2+6\rho_1\rho_2 \right) + 8\left(\omega \lambda_1 \rho_1+\lambda_2\rho_2 \right)},\\
Z^2= ~&\frac{18 \left[\kappa_2 \lambda_1 \lambda_2 \left(2\kappa_1+1\right) -4\left(\omega \lambda_1 \rho_1 +\lambda_2 \rho_2 \right) \right] } {\left[\lambda_1\lambda_2 \left(2\kappa_1+1\right) \left(\kappa_2+6\rho_1\rho_2 \right) + 8\left(\omega \lambda_1 \rho_1+\lambda_2\rho_2 \right)\right]^2} \nonumber\\
& \times \left\{\lambda_1 \lambda_2 \left[ 2\rho_1 \rho_2 \left(6\kappa_1+1\right) +\kappa_2 \left(2\kappa_1-1\right) \right] +8\left(\omega \lambda_1 \rho_1 +\lambda_2 \rho_2 \right)\right\} .
\end{align}
It is noted that a trivial solution corresponding to $Z^2=0$ has been ignored. 
It is also noted that the corresponding value of $Y_2$ can be obtained in terms  of $Y_1$ as shown in Eq. \eqref{relation-Y}. It is straightforward to see that this anisotropic fixed point is equivalent to the anisotropic power-law solutions found in the previous section. Indeed, one can easily verify that $X=\eta/\zeta$, according to the solution shown in Eqs.  \eqref{solution-zeta}, \eqref{solution-eta}, and \eqref{solution-X}. 

As a result, during the inflationary phase with $\rho_i \gg \lambda_i$, we can approximate the anisotropic fixed point as
\begin{align}
X& \simeq \frac{1}{3\kappa_1} \ll 1,\\
Y_1& \simeq -\frac{2}{\rho_1}\ll1,\\
Y_2 &\simeq -\frac{2}{\rho_2 }\ll 1,\\
Z^2 &\simeq 9X \ll 1, \\
W_1^2+W_2^2& \simeq 3.
\end{align}
In addition, it turns out that $Z^2 \ll 1$ implies that $Z_1^2 \ll1$ as well as $Z_2^2 \ll1$, according to the definition in Eq. \eqref{definition-of-Z}. 

\begin{figure}[hbtp] 
\begin{center}
{\includegraphics[height=85mm]{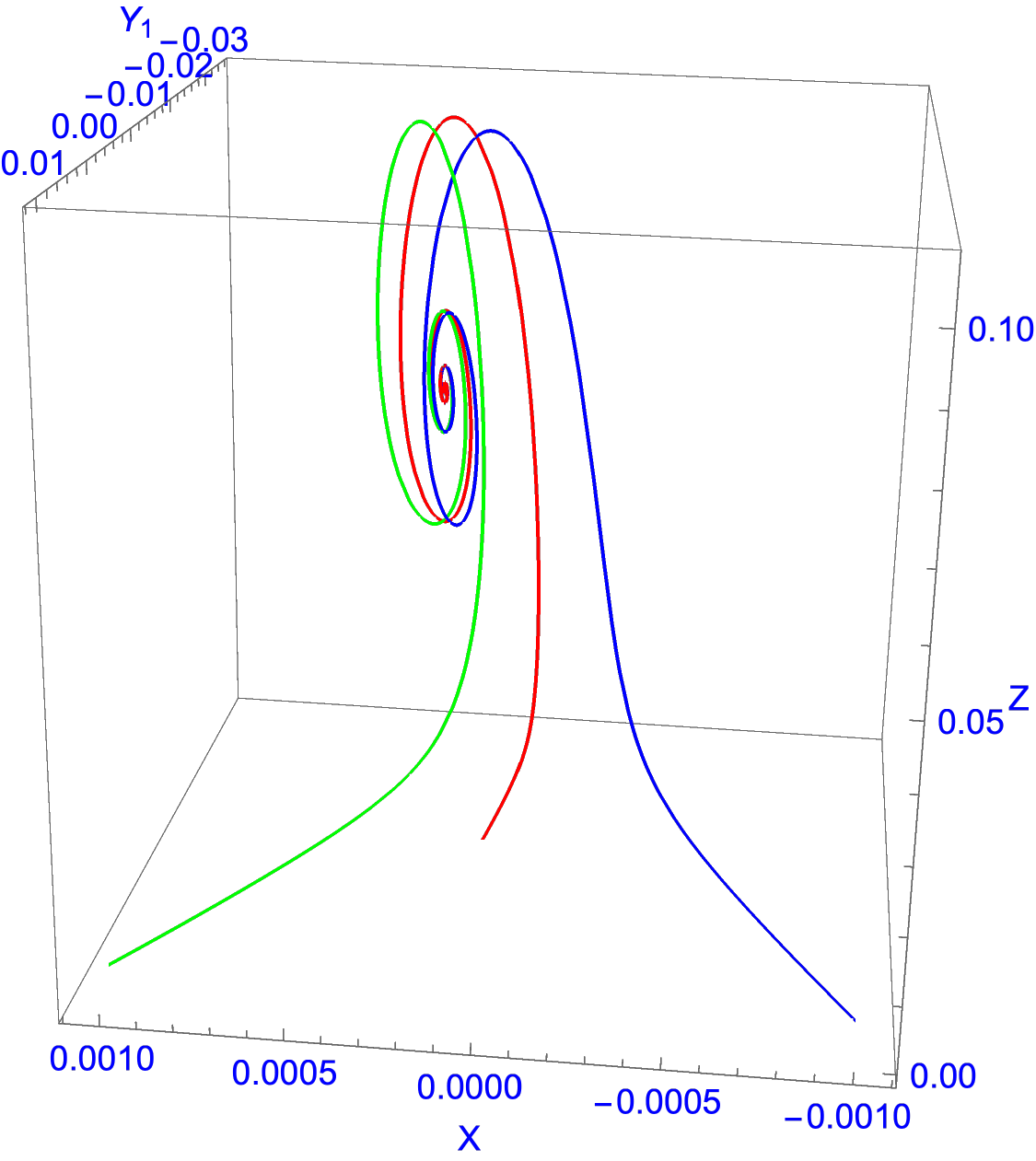}}\quad
{\includegraphics[height=85mm]{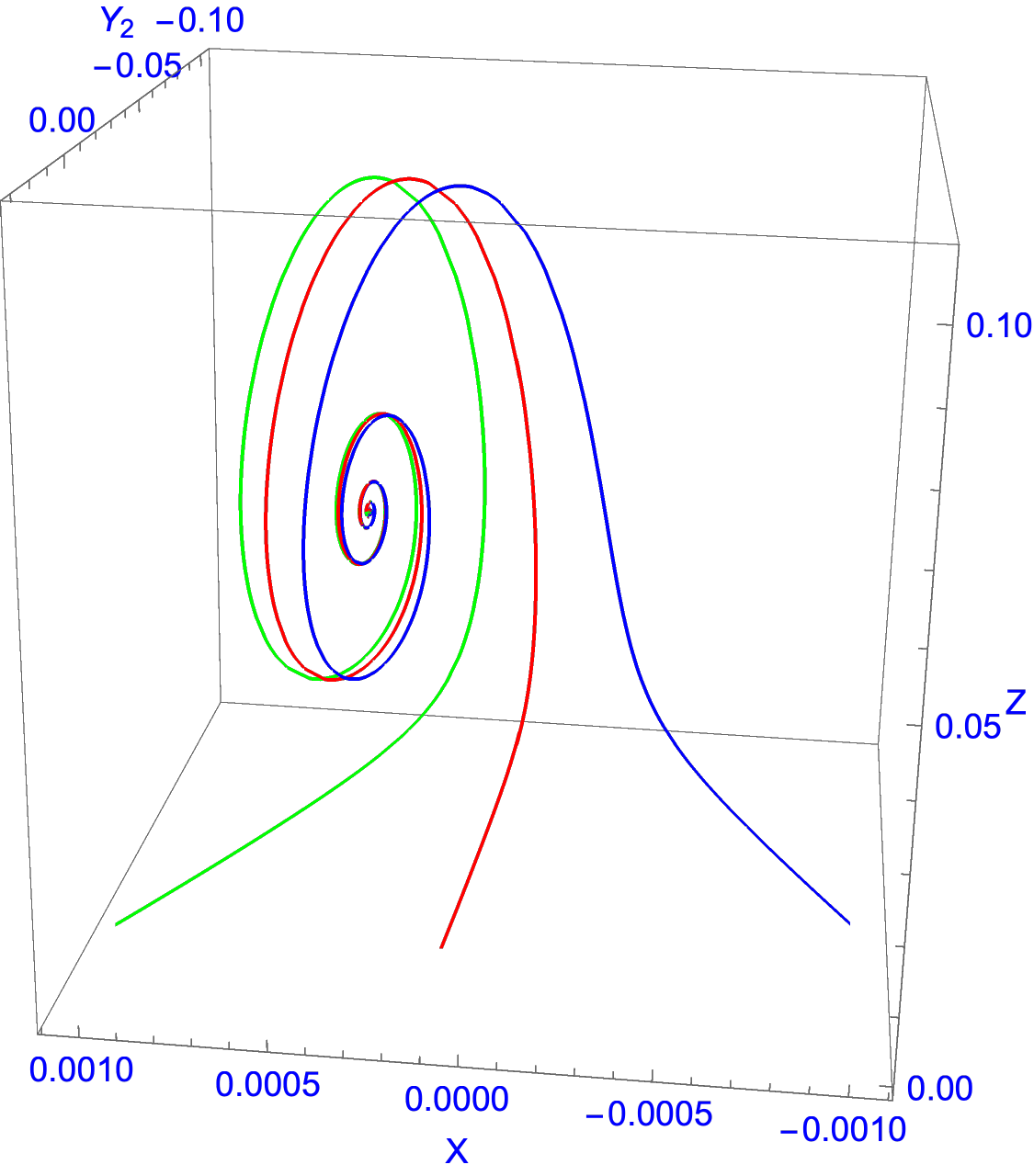}}\\
\caption{Attractor behavior of the anisotropic fixed point with the field parameters chosen as $\omega=+1$, $\lambda_1=0.1$, $\lambda_2=0.2$, $\rho_1 =50$, and $\rho_2=100$. These two plots clearly display that trajectories with different colors corresponding to different initial conditions all converge to the anisotropic fixed point.}
\label{fig1}
\end{center}
\end{figure}

Now, we would like to investigate the stability of the obtained anisotropic fixed point by perturbing the dynamical system around this fixed point as follows \cite{MW,Do:2017rva}
\begin{align}
\frac{d\delta X}{d\alpha} &\simeq -3\delta X,\\
\frac{d\delta Y_1}{d\alpha} &\simeq -3\delta Y_1 +2\rho_1 Z_1\delta Z_1 -2\lambda_1 W_1 \delta W_1,\\
\frac{d\delta Y_2}{d\alpha} &\simeq -3\delta Y_2 + \frac{2\rho_2}{\omega}  Z_2\delta Z_2 - \frac{2\lambda_2}{\omega}  W_2 \delta W_2,\\
\frac{d\delta Z_1}{d\alpha}&\simeq -Z_1 \left(2\delta X+\rho_1 \delta Y_1\right),\\
\frac{d\delta Z_2}{d\alpha}&\simeq -Z_2 \left(2\delta X+\rho_2 \delta Y_2\right),\\
\frac{d\delta W_1}{d\alpha}&\simeq \frac{\lambda_1}{2}W_1 \delta Y_1,\\
\frac{d\delta W_2}{d\alpha}&\simeq \frac{\lambda_2}{2}W_2 \delta Y_2.
\end{align}
Taking the exponential perturbations as
\begin{align}
&\delta X =A_1 \exp[\tau \alpha];~\delta Y_1 =A_{2} \exp[\tau \alpha];~\delta Y_2 =A_{3} \exp[\tau \alpha], \nonumber\\
&\delta Z_1 =A_{4} \exp[\tau \alpha];~\delta Z_2 =A_{5} \exp[\tau\alpha];~\delta W_1 =A_{6} \exp[\tau \alpha];~\delta W_2 =A_{7} \exp[\tau \alpha],
\end{align}
will lead the above perturbed equations to the following homogeneous linear system of $A_i$, which can be written as a homogeneous matrix equation as
\begin{equation} \label{stability-equation}
{\cal M}\left( {\begin{array}{*{20}c}
   A_1  \\
   A_{2}  \\
   A_{3} \\
   A_{4}  \\
   A_{5}  \\
   A_{6}\\
   A_{7}\\
 \end{array} } \right) \equiv \left[ {\begin{array}{*{20}c}
   {-3-\tau} & {0} & {0 } & {0 } & {0} &{0} &{0} \\
   {0 } & {-3-\tau} & {0 } & {2\rho_1 Z_1} &{0}&{-2\lambda_1 W_1}&{0}  \\
     {0 } & {0} & {-3-\tau } & {0 } &{\frac{2\rho_2}{\omega}  Z_2}&{0}&{-\frac{2\lambda_2}{\omega} W_2}  \\
   {-2Z_1} & {-\rho_1 Z_1 } & {0 } & {-\tau } &{0}&{0}&{0} \\
   {-2Z_2} & {0 } & {-\rho_2 Z_2 } &{0}& {-\tau } &{0}&{0}\\
   {0}&{\frac{\lambda_1 }{2}W_1}&{0} &{0} &{0}&{-\tau}&{0}\\
{0}&{0}&{\frac{\lambda_2 }{2}W_2} &{0} &{0}&{0}&{-\tau}\\
 \end{array} } \right]\left( {\begin{array}{*{20}c}
    A_1  \\
   A_{2}  \\
   A_{3} \\
   A_{4}  \\
   A_{5}  \\
   A_{6}\\
   A_{7}\\
 \end{array} } \right) = 0.
\end{equation}
Mathematically, non-trivial solutions, i.e., $A_i \neq 0$, of the homogeneous linear system exist if and only if
\begin{equation}
\det {\cal M}=0,
\end{equation}
which can be reduced to the following equation of $\tau$,
\begin{equation} \label{equation-of-tau}
\tau^2 \left(\tau+3\right) \left(\tau^2 +3\tau +\lambda_1^2W_1^2 +2\rho_1^2 Z_1^2 \right)\left(\omega\tau^2 +3\omega\tau +\lambda_2^2W_2^2 +2\rho_2^2 Z_2^2 \right)=0.
\end{equation}
Now, we would like to examine whether the equation, $\det {\cal M}=0$, admits any positive root $\tau>0$ corresponding to unstable modes. For $\omega =+1$, it is straightforward to see that Eq. \eqref{equation-of-tau}  does not admit any positive root $\tau>0$ since its coefficients all  turn out to be positive definite. Therefore, the corresponding anisotropic fixed point  turns out to be stable during the inflationary phase. On the other hand, it appears for $\omega =-1$ that Eq. \eqref{equation-of-tau} does admit  at least one positive root $\tau>0$, which is nothing but that of the equation, 
\begin{equation}
-\tau^2 -3\tau +\lambda_2^2W_2^2 +2\rho_2^2 Z_2^2 =0.
\end{equation}
This result implies that the corresponding anisotropic fixed point is indeed  unstable during the inflationary phase. This result is consistent with our previous investigations in Refs. \cite{WFK,Do:2017rva,non-canonical}, in which we have shown that the inclusion of the phantom field with $\omega=-1$ breaks down the stability of the anisotropic inflation. It is worth noting that the stability of the anisotropic fixed point can be numerically confirmed through its attractor behavior. In particular, the stable or unstable fixed points will be shown to be attractive or unattractive, respectively. Therefore, we would like to examine whether this anisotropic fixed point is an attractor one or not, similar to the previous studies \cite{MW,non-canonical,Do:2018zac}. To do this, we will numerically solve the dynamical system with different initial conditions and then plot the corresponding phase spaces of dynamical variables $X$, $Y_1$, $Y_2$, and $Z$. As a result, the numerical plots in Fig. \ref{fig1} confirm that the anisotropic fixed point is indeed attractive for $\omega=+1$ as expected. For $\omega=-1$, it turns out that the anisotropic fixed point is unattractive as expected since all  trajectories converge to the isotropic fixed point with $X=Z=0$. 
\section{Conclusions} \label{sec4}
We have proposed a multi-field extension of the KSW anisotropic inflation, in which two scalar fields, $\phi$ and $\psi$, are allowed to non-minimally couple to two vector fields, $A_\mu$ and ${\cal A}_\mu$, respectively, through two corresponding couplings, $f_1^2(\phi)F_{\mu\nu}F^{\mu\nu}$ and $f_2^2(\psi) {\cal F}_{\mu\nu} {\cal F}^{\mu\nu} $.  As a result, we have found an exact anisotropic power-law solution to this model. It turns out that we always have the result $\zeta \simeq \kappa_1$ for anisotropic power-law solutions, regardless of the nature of scalar fields as well as the number of scalar and vector fields counted in the KSW model. However, it appears that the $\zeta$ solution, $\zeta \simeq \kappa_1$,  is a half of that obtained in the two-scalar-field models \cite{WFK,Do:2017rva}; while the $\eta$ solution, $\eta \simeq 1/3$, remains the same, assuming $\rho_1/\lambda_1 =\rho_2/\lambda_2 \gg 1$.
More interestingly, this solution has been shown in the case, in which both scalar fields are canonical, through a dynamical system method, to be stable and attractive during the inflationary phase. Hence, the cosmic no-hair conjecture is really violated in this case. However, the stability of the found solution has been shown to be broken down once one of these two scalar fields is phantom with $\omega=-1$. This result together with previous investigations \cite{WFK} indicate that the phantom field does favor the conjecture. It should be noted that this paper is devoted to examine the validity of the cosmic no-hair conjecture within the multi-field extension of the KSW anisotropic inflation. Other cosmological aspects such as the CMB imprints \cite{Imprint1,Imprint2,Imprint3,Do:2020ler} of this model will be our future studies and will be presented elsewhere.  We hope that our model would be useful to studies of the early time universe. 
\begin{acknowledgments}
T.Q.D. is supported by the Vietnam National Foundation for Science and Technology Development (NAFOSTED) under grant number 103.01-2020.15. W.F.K. is supported in part by the Ministry of Science and Technology (MOST) of Taiwan under Contract No. MOST 109-2112-M-009-001. 
\end{acknowledgments}

\end{document}